\documentclass[longbibliography,showpacs,floatfix,nofootinbib,superscriptaddress,twocolumn,aps,prb]{revtex4-2}

\usepackage{graphicx,color}
\usepackage{amsfonts}
\usepackage[figuresright]{rotating}  
\usepackage{amssymb}
\usepackage{bm}
\usepackage{amsmath}
\usepackage{mathtools}
\usepackage{psfrag}
\usepackage{multirow}
\usepackage{tabularx}
\usepackage{textcomp}
\usepackage{units}
\usepackage{lipsum}
\usepackage{wasysym} 
\usepackage{graphicx}
\usepackage{physics}
\usepackage{soul}
\usepackage{times}
\usepackage[colorlinks=true,citecolor=blue,linkcolor=magenta,hypertexnames=false]{hyperref}
\usepackage{enumitem}
\usepackage{floatrow}
\usepackage[dvipsnames]{xcolor}
\usepackage[caption=false]{subfig}
\usepackage{tikz}
\usepackage{appendix}

\renewcommand{\selectlanguage}[1]{}

\begin{document}

\title{Fragile spin liquid in three dimensions}

\author{Anna Fancelli$^*$}
\affiliation{Dahlem Center for Complex Quantum Systems and Fachbereich Physik, Freie Universität Berlin, Arnimallee 14, 14195 Berlin, Germany}
\affiliation{Helmholtz-Zentrum Berlin für Materialien und Energie GmbH, Hahn-Meitner-Platz 1
14109 Berlin, Germany}

\author{R. Flores-Calderón$^*$}
\affiliation{Max Planck Institute for the Physics of Complex Systems, Nöthnitzer Strasse 38, 01187 Dresden, Germany}
\affiliation{Max Planck Institute for Chemical Physics of Solids, Nöthnitzer Strasse 40, 01187 Dresden, Germany}

\author{Owen Benton}
\affiliation{School of Physical and Chemical Sciences, Queen Mary University of London, London, E1 4NS, United Kingdom}

\author{Bella Lake}
\affiliation{Helmholtz-Zentrum Berlin für Materialien und Energie GmbH, Hahn-Meitner-Platz 1
14109 Berlin, Germany}
\affiliation{Institut f\"ur Festk\"orperforschung, Technische Universit\"at Berlin, 10623 Berlin, Germany}

\author{Roderich Moessner}
\affiliation{Max Planck Institute for the Physics of Complex Systems, Nöthnitzer Strasse 38, 01187 Dresden, Germany}

\author{Johannes Reuther}
\affiliation{Dahlem Center for Complex Quantum Systems and Fachbereich Physik,
Freie Universität Berlin, Arnimallee 14, 14195 Berlin, Germany}

\affiliation{Helmholtz-Zentrum Berlin für Materialien und Energie GmbH, Hahn-Meitner-Platz 1
14109 Berlin, Germany}

\begin{abstract}
Motivated by the recent appearance of the trillium lattice in the search for materials hosting spin liquids, we study the ground state of the classical Heisenberg model on its linegraph, {the trilline lattice}. We find that this network realises the recently proposed notion of a fragile spin liquid in three dimensions. Additionally, we analyze the Ising case and argue for a possible $\mathbb{Z}_2$ quantum spin liquid phase in the corresponding quantum dimer model. Like the well-known $U(1)$ spin liquids, the classical phase hosts moment fractionalisation evidenced in the diluted lattice, but unlike these, it exhibits exponential decay both in spin correlations and interactions between fractionalised moments. This provides the first instance of a purely short-range correlated classical Heisenberg spin liquid in three dimensions. 
 \end{abstract}
\maketitle

\def\thefootnote{*}\footnotetext{These authors contributed equally to this work}\def\thefootnote{\arabic{footnote}}

\section{Introduction}

The interplay between frustration, topology, and strong interactions has
opened a window into the limits of what matter can be and do. Because of
their exotic properties, spin liquids represent natural settings for
such explorations \cite{anderson_resonating_1973,balents_spin_2010,zhou_quantum_2017,wen_quantum_2002,castelnovo_spin_2012,FieldGuide}. Recently much work has been done on the forefront of generalizing the low energy theory from conventional gauge theories to higher rank gauge theories \cite{xu_gapless_2006,chamon_quantum_2005,haah_local_2011,vijay_fracton_2016,vijay_new_2015,halasz_fracton_2017,you_building_2019,xu_bond_2007,xu_emergent_2010,benton_spin-liquid_2016,benton_topological_2021,pretko_generalized_2017,yan_rank--2_2020,flores-calderon_irrational_2024}. An example is the recently theorized rank-2 $U(1)$ spin liquid, which is described by a traceless symmetric tensor instead of the usual vector potential of electromagnetism. These exotic extensions to our fundamental theories are not realized in our universe, so their study and realization may unravel physical effects that cannot otherwise be studied.

On another front, a recent classification scheme for classical spin liquids (CSLs) has been put forward and the notion of a fragile spin liquid has emerged \cite{yan_classification1_2023,yan_classification2_2023}. This type of spin liquid is a phase of matter presenting the phenomena of fractionalization, as is the case for other algebraic spin liquids, while retaining an energy gap associated with exponentially decaying correlations. The idea of such a short-range correlated state is not new.  We may recall the quintessential example for a strongly correlated topological phase of matter, the toric code. Originally defined in the square lattice, this model presents fractionalized excitations in the form of emergent electric and magnetic particles with semionic statistics \cite{Kitaev2006,Kitaevmemory,Wen1991,StringNet}. The exactly known ground state has a gap to excited states realizing the celebrated $\mathbb{Z}_2$ topological order. In a similar spirit we present here the first instance of a 3D fragile spin liquid, which acts as the classical analog to the $\mathbb{Z}_2$ quantum spin liquids.
 
The theoretical identification of novel spin liquids is also of high experimental relevance, given the lack of conclusive physical spin liquid realizations and the quest to identify experimentally relevant signatures. Recently, it has been proposed that a classical Heisenberg model on the distorted windmill lattice, inspired by the spin-$1/2$  spin liquid candidate $\text{PbCuTe}_2\text{O}_6$ \cite{Chillal2020Evidence}, exhibits an extensive ground state degeneracy characteristic of classical spin liquid behaviour~\cite{windmillSLs}. For a particular arrangement of spin couplings this CSL can be understood as a spin model on the line graph of the trillium lattice, which we name the \textit{trilline lattice}. While the trillium lattice consists of a non-centrosymmetric network of corner sharing equilateral triangles the trilline lattice consists of distorted corner-sharing octahedra. Of particular relevance is its characteristic non-bipartite structure, which is known in 2D to change the nature of the frustrated magnetic phase. We may compare this situation to the well known case of spin ice, described by the pyrochlore lattice, a network of corner sharing tetrahedra, which permits a bipartite lattice description. This property together with the spin ice constraint of vanishing net spin on each tetrahedron ultimately leads to its remarkable $U(1)$ gauge structure. 

In contrast, previous works have noted that for Ising spins in a non-bipartite lattice, sectors of the configuration space are labelled by $\mathbb{Z}_2$ winding parities rather than the $\mathbb{Z}$-valued winding numbers of $U(1)$ gauge theories realized on bipartite lattices \cite{rehn_fractionalized_2017}. When translated to the disordered dimer coverings on non-bipartite lattices it was found that the entropic interaction potential between a pair of monomers approaches a finite limiting value exponentially fast with increasing separation \cite{DimerModelLiquids,triangdimer,qdmKagome}. The deconfinement of monomers is a telltale sign of fractionalization; however,   dimer-based descriptions are natural  for discrete Ising spins, while the dilution signatures studied in this paper can be more generally applied to other, in particular continuous, spin models \cite{sen_fractional_2011,rehn_classical_2016,sen_vacancy-induced_2012} and even to the quantum versions of these models \cite{quantumorphans}.

The rest of the paper is structured as follows. In Sec.~\ref{sec:model} we describe our spin model on the trilline lattice and its relation to previously studied frustrated models. Next, in Sec.~\ref{sec:characterization}, we characterize the spin correlations via the structure factor in momentum space and analyze how our model fits in the recently introduced classification of CSLs \cite{yan_classification1_2023,yan_classification2_2023}. In Sec.~\ref{sec:dilution} we probe the fractionalization properties of the system via the introduction of non-magnetic vacancies, and we
find that the most relevant dilution clusters behave as free spins with a fractional magnetic
moment of $S/2$. Together with exponentially decaying correlations
between the dilution clusters, the analog of gauge charges, this distinguishes gapped 3D $\mathbb{Z}_2$
spin liquids from algebraic ones, and positions them as an
instance of fractionalization with purely short range
correlations. Finally, in Sec.~\ref{Sec:IsingandQuant} we discuss the Ising version of our spin model on the trilline lattice which we rewrite as a dimer model and elaborate on the effects of dimer resonances. We argue that such resonances can be tuned to a point of genuine {\it quantum} spin liquid behavior, known as the RK point. The paper ends with a conclusion in Sec.~\ref{sec:conclusion}.

 \begin{figure}[ht!]
     \centering
     \includegraphics[width=1\columnwidth]{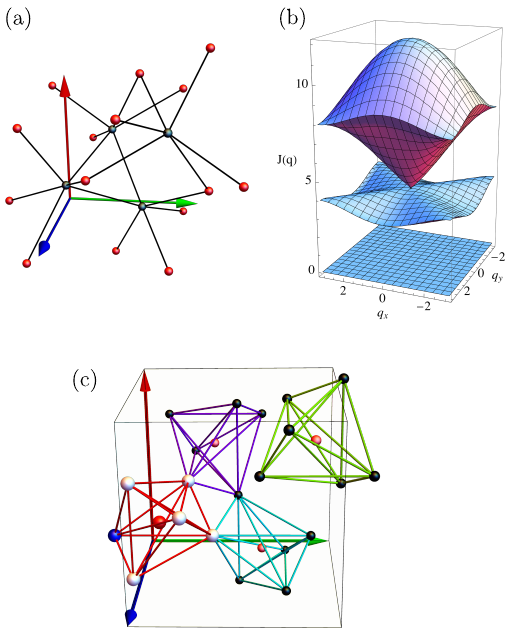}
     \caption{(a) Trillium lattice, with the four unit cell sites shown in gray and the corresponding nearest neighbour sites shown in red. We note that the site positions of trillium lattice depend on one real parameter. Here and in the following, we fix this parameter to the value realized in $\text{PbCuTe}_2\text{O}_6$, see Ref.~\cite{windmillSLs} for details. (b) Dispersion of the interaction matrix eigenvalues in SCGA  in the $q_z=0$ plane. (c) Unit cell with the four inequivalent octahedra of the trilline lattice in distinct colors. A collection of vacancy spins (white) yields an orphan spin (blue). The centers of the octahedra (red spheres) correspond to the trillium lattice.}
     \label{fig:Jq_orph}
 \end{figure}
\section{Model}\label{sec:model}
We consider classical $O(3)$ Heisenberg spins in the line-graph of the trillium lattice with the Hamiltonian:
\begin{align}
    \mathcal{H}= \dfrac{J}{2}\sum_{\mathrm{oct}}\left(\sum_{i \in \mathrm{oct}} \vec{S}_i\right)^2,\label{Ham}
\end{align}
where $\mathrm{oct}$ refers to the distorted octahedral cages of the line-graph of the trillium lattice [Fig \ref{fig:Jq_orph}(a)], what we call the trilline lattice. The trillium lattice \cite{Hopkinson2006trillium} from which the trilline lattice is constructed is a lattice with a four-atomic cubic unit cell consisting of corner-sharing equilateral triangles, where each lattice site is part of three triangles. A line-graph is constructed by considering the midpoints of all the nearest neighbours bonds for each distinct point in the original graph, here forming the distorted octahedral cages, see Fig. \ref{fig:Jq_orph}(c). As also illustrated in Fig. \ref{fig:Jq_orph}(c) the centers of the distorted octahedra are the sites of the trillium lattice and thus the trillium lattice is the dual to the non-bipartite trilline lattice, when considering the dual graph as built from the centers of the elementary 2-chains. Below we investigate the properties of the model (and a diluted version thereof) in a large $\mathcal{N}$ approximation, also known as the self-consistent Gaussian approximation (SCGA) \cite{garanin_classical_1999,IsakovSohndiSpinIce,sen_vacancy-induced_2012,rehn_classical_2016,IsakovSondhiLargeN}, where the number of spin components is generalized from three to $\mathcal{N}$, see Appendix~\ref{app_a} for details on the method. Within this formalism we consider the interaction matrix $J(\textbf{q})$ in momentum space which allows us to rewrite the Hamiltonian in Eq.~\eqref{Ham} as:
\begin{align}
    \mathcal{H}=\frac{1}{2} \sum_{\mathbf{q}} \sum_{a, b=1}^n \tilde{S}_a(-\mathbf{q}) J_{a b}(\mathbf{q}) \tilde{S}_b(\mathbf{q}),
\end{align}
where we have defined the Fourier transformed spin field $\tilde{S}_b(\mathbf{q})$, and $n=12$ is the number of sublattices in the trilline lattice. Diagonalizing the matrix $J_{a b}(\mathbf{q}) $ results in a spectrum which contains eight degenerate bottom flat bands and four dispersing ones with a linearly dispersing cone at finite energy and zero momentum separated by a gap, as shown in Fig.~\ref{fig:Jq_orph}(b). This band-touching point is a Weyl point which has a corresponding partner at the Brillouin zone corner $(\pi,\pi,\pi)$, consistent with the fact that only an even number of Weyl points is allowed in lattice models. This feature is interesting in itself, as it suggests a non-trivial source of Berry curvature from each Weyl point, which may have an analogue to topological Weyl magnons, of much interest to the magnetism and transport community \cite{Weylmagnons2016}. 

It is worth noting that the ground state flat bands are of crucial importance for the existence of a spin liquid phase. Physically, such zero-modes represent the allowed transformations which keep the ground state constraint, $\sum_{i \in \mathrm{oct}} \vec{S}_i=0$, implied by Eq.~\eqref{Ham} intact. In contrast, the upper dispersive bands represent the configurations which violate the ground state constraint. These excited states are usually not in a simple relationship to the flat band states. However, in algebraic spin liquids like the $U(1)$ spin ice, the excited states of the dispersive bands can be understood as the charges of the emergent gauge theory added to the ground states. 

 \begin{figure}[ht]
     \centering
     \includegraphics[width=1\columnwidth]{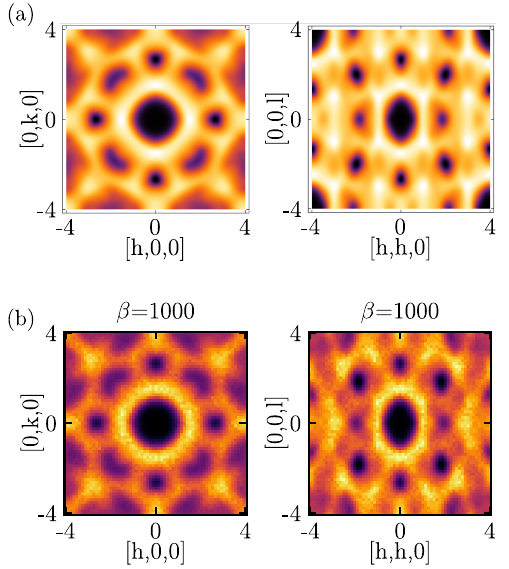}
     \caption{Static spin structure factor $\mathcal{S}(\textbf{q})$ calculated within (a) the SCGA scheme for the $(h,0,0)$ vs. $(0,k,0)$ and for the $(h,h,0)$ vs. $(0,0,l)$ planes. (b) shows the analogous plots for the Heisenberg case obtained from MC simulations at the lowest simulated temperature, corresponding to $\beta=1000$.}
     \label{fig:SSF}
 \end{figure}
\section{Characterization of the spin liquid phase}\label{sec:characterization}
To analyze the ground state phase of our model we consider the static structure factor defined as:
\begin{align}
    \mathcal{S}(\textbf{q})=\expval{\vec{S}(\textbf{q})\cdot \vec{S}(-\textbf{q}))}.
\end{align}
If the system were magnetically ordered, we would see the presence of sharp Bragg peaks at specific momentum points. Instead, we see only smooth, broad features as exemplified in Fig.~\ref{fig:SSF}, which displays cuts of the 3D Brillouin zone along two planes. Within the SCGA [see Fig.~\ref{fig:SSF}(a)] we may compute the spin structure factor for small enough temperatures simply as the projector to the flat bands $\mathcal{S}(\textbf{q})=\mathcal{P}_{\text{flat}}(\textbf{q})$ which in our case are eightfold degenerate. The absence of Bragg peaks indicates no ordering to the lowest temperatures while the absence of pinch-point singularities, so quintessential of spin ice \cite{moessner_pyrochlore_1998,moessner_low-temperature_1998,moessner_magnetic_1999,IsakovSohndiSpinIce,andre_frustration_1979,garanin_self-consistent_1996}, further confirms the short-range exponentially decaying correlations of the spins. We simulate the three-component ($\mathcal{N}=3$) Heisenberg system using classical Monte Carlo (MC) methods (see Appendix~\ref{app_b} for details). Throughout, energy and temperature are expressed in units of $J$, with $J=1$. The spin structure factor obtained from MC simulations, shown in Fig.~\ref{fig:SSF}(b), exhibits a qualitatively similar signal distribution to the one obtained from SGCA in Fig.~\ref{fig:SSF}(a), with no peaks or other singularities.

Further characterization of the spin liquid phase can be obtained by studying the constraint of the Hamiltonian by the methods and classification described in detail in Ref.~\cite{yan_classification1_2023,yan_classification2_2023}, which we follow next. Our goal will be to identify to which class of momentum-space mappings our model belongs within this classification scheme. We conjecture and later show by means of dilution that our model belongs to a fractionalized phase. Let us start by defining the constrainer, which is a sum of spins that is zero in the ground state, thus constraining the possible ground state spin configurations. We note first that the unit cell has four octahedra as shown in Fig.~\ref{fig:Jq_orph}(a), which means that we have $m=4$ constrainers and $n=12$ sublattices implying $n-m=8$ flat bands as mentioned previously. We write the Hamiltonian in the constrainer form:
\begin{align}
\mathcal{H} & =\dfrac{J}{2}\sum_{\mathbf{R} \in \text { u.c. }} \sum_{I=1}^m\left[\mathcal{C}_I(\mathbf{R})\right]^2\notag \\
& =\dfrac{J}{2}\sum_{\mathbf{R} \in \text { u.c. }} \sum_{I=1}^m\left[\sum_{\mathbf{r}} \mathbf{S}(\mathbf{r}) \cdot \mathbf{C}_I(\mathbf{R}, \mathbf{r})\right]^2.\label{eq:constrainer}
\end{align}
In the first line, $\mathcal{C}_I(\mathbf{R})$ denotes the constrainers of our model. For a unit cell (u.c.) at position $\mathbf{R}$ and for an octahedron $I$ within that unit cell the constrainer $\mathcal{C}_I(\mathbf{R})=\sum_{i\in\text{oct}(\mathbf{R},I)}S_{i}$~\cite{footnote} is the sum of spins in this octahedron, indicated by the notation $i\in\text{oct}(\mathbf{R},I)$. In the second line of Eq.~(\ref{eq:constrainer}) this is rewritten using a vector notation over sublattices indicates $\alpha=1,\dots,12$ where $\mathbf{S}(\mathbf{r})= (S_1(\mathbf{r}),\dots,S_{12}(\mathbf{r}))$ and $\mathbf{r}$ denotes the positions of the spins. The entries of $\mathbf{C}_I(\mathbf{R}, \mathbf{r})$ are $C^\alpha_I(\mathbf{R}, \mathbf{r})=1$ if the octahedron $I$ in the unit cell $\mathbf{R}$ has a site on sublattice $\alpha$ and if $\mathbf{r}$ matches the position of that site. In all other cases $C^\alpha_I(\mathbf{R}, \mathbf{r})=0$. By Fourier transforming we obtain the so called $FT$-constrainers denoted by $\textbf{T}_I(\textbf{q})= \sum_{\textbf{r}}\mathbf{C}_I(\mathbf{R}, \mathbf{r}) e^{i\textbf{q}\cdot \textbf{r}} $. The $FT$-constrainers span the eigenspace of all non-flat bands; in our case we have $m=4$ of these bands. We further identify vectors $\textbf{T}_I(\textbf{q})$ which differ by a complex number, $\textbf{T}_I(\textbf{q})\sim c\, \textbf{T}_I(\textbf{q})$ with $c\in\mathbb{C}$ since they correspond to the same spin configurations. This reduces the $n=12$ dimensional complex vector space of the $FT$-constrainers to the complex projective space $\mathbb{C}P^{n-1}$.  

Using the fact that the topology of the Brillouin zone is that of a three-torus, we focus on mappings from $T^3$ to a subset of $\mathbb{C}P^{11}$ generated by $\textbf{T}_1(\textbf{q}),\textbf{T}_2(\textbf{q}),\textbf{T}_3(\textbf{q}),\textbf{T}_4(\textbf{q})$. The space of all $m$-dimensional subspaces in $\mathbb{C}P^{n-1}$ is known as the complex Grassmanian, denoted by $\text{Gr}(m,n)$. For a single dispersive band $\text{Gr}(1,n)=\mathbb{C}P^{n-1}$ as expected \cite{yan_classification1_2023}. In our case $m=4,n=12$ so the appropriate space to consider is $\text{Gr}(4,12)$. Since calculating the homotopy group $[T^3,\text{Gr}(4,12)]$ and determining if our model falls into a topologically non-trivial class is an arduous task, we choose instead to focus on the physically relevant signatures that we would expect in such a case. If we find an observable that takes a distinctly different value than for a trivial paramagnet we can conjecture that our system lies in a topologically non-trivial sector. In the following, we identify such an observable as the emergent magnetic moment of certain dilution clusters.

 \begin{figure}[b]
     \centering
     \includegraphics[width=1\columnwidth]{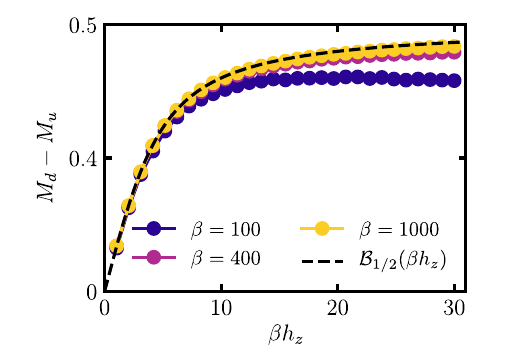}
     \caption{Magnetisation of the diluted system $M_d$ (one orphan spin), minus that of the undiluted case $M_u$, as a function of inverse temperature and magnetic field from classical MC. We fix the temperature and scan the magnetic field. The black-dashed line shows the theoretical value of a fractionalised moment of one half, given by the Brillouin function $\mathcal{B}_s(h_z \beta)$ for $S=1/2$. This function describes the magnetization of a paramagnet with magnetic moment $S$ in response to an external magnetic field. }
     \label{fig:Mag}
 \end{figure}
 
\section{Dilution and fractionalization}\label{sec:dilution}

To further probe the emergent cooperative behaviour of our system, we study the response to non-magnetic vacancies in the form of dilution.
The simplest case is that of a single spin being removed from a given octahedron. Starting from the ground state of the Hamiltonian in Eq.~(\ref{Ham}) we observe that the total magnetization in each distorted octahedron must be zero. The effect of removing a single spin can then be compensated by rearranging the remaining spins such that the local ground state constraints are still fulfilled. Such a smooth deformation is allowed by the continuous $O(3)$ nature of the spins. Hence, we do not expect a single missing spin to significantly change the bulk response.

This argument can be extended to multiple spins being removed from a single cluster until all but one of its spins have been removed. Such a configuration implies $\vec{S}_{\text{oct}}=\sum_{i\in\text{oct}}\vec{S}_{i}\equiv\vec{S}_{\text{O}}\neq 0$, where $\vec{S}_{\text{O}}$ is the last remaining \textit{orphan spin} in this octahedron. An example for such a dilution configuration is shown in Fig.~\ref{fig:Jq_orph}(c), where we illustrate the missing sites by white spheres, the centers of the distorted octahedra in red and draw the bonds in the different octahedra in different colors. Furthermore, the orphan spin is shown in blue. Because of the hard spin length constraint, there is no possibility of smoothly compensating such dilution. An orphan spin cluster will thus dominate the magnetic response at low temperatures. This can be seen from its free-spin behaviour but with a fractional spin magnitude that implies a diverging susceptibility contribution at zero temperature, as noted in Ref.~\cite{sen_fractional_2011}.

Moreover, we can understand these dilution clusters from the perspective of low energy theory. In the usual case of $U(1)$ spin liquids such configurations violate the local gauge constraint, and we can think of them as chemically placed gauge charges. In fact, it is known that such configurations probe the emergent fractionalization properties of other spin liquids as well \cite{flores-calderon_irrational_2024,sen_fractional_2011,rehn_random_2015,rehn_fractionalized_2017}. We will show that this is also the case for our fragile spin liquid in 3D. 

\subsection{Fractional magnetic moment}
The trilline lattice is composed of corner sharing distorted octahedra. Crucially this implies that we can write the Hamiltonian in a magnetic field $\vec{h}$ as:
\begin{align}
    \mathcal{H}&= \dfrac{J}{2}\sum_{\mathrm{oct}}\left(\sum_{i \in \mathrm{oct}} \vec{S}_i\right)^2-\sum_i \vec{h}\cdot\vec{S}_i\\
    &=\dfrac{J}{2}\sum_{\mathrm{oct}}\left(\sum_{i \in \mathrm{oct}} \vec{S}_i\right)^2-\dfrac{1}{2}\sum_{\mathrm{oct}}\sum_{i \in \mathrm{oct}}  \vec{h}\cdot\vec{S}_i\\
    &=\dfrac{J}{2}\sum_{\mathrm{oct}}\left(\sum_{i \in \mathrm{oct}} \vec{S}_i-\dfrac{\vec{h}}{2J}\right)^2+\text{const.}\;.\label{Horph}
\end{align}
Since an orphan spin in a specific octahedron $\mathrm{oct}$ implies $\sum_{i\in\text{oct}}\vec{S}_{i}\equiv\vec{S}_{\text{O}}$, expanding the first term in Eq.~\eqref{Horph} yields a contribution from this octahedron $\mathcal{H}_{\text{O}}= -\dfrac{1}{2} \vec{h}\cdot\vec{S}_{\text{O}}$. This can be read as a reduced magnetic moment of a free spin of magnitude $\alpha S = S/2$. This argument assumes a certain independence between the spin clusters and could still be interpreted as a reduction of the magnetic field instead of a change of magnetic moment. Nevertheless, from our MC simulations and the more involved vacancy field theory (FT), we find that the behaviour, for small fields and low temperatures, is indeed quantitatively described by a free $S/2$ magnetic moment, as shown in Fig.~\ref{fig:Mag}. There, we plot the diluted minus the undiluted magnetization from MC as a function of $\beta h_z$ for fixed temperatures.

As a final note, the magnetic moment found here fractionalizes with the specific value $S/2$. It has been shown that the magnetic moment of such an orphan spin can be tuned continuously and even assume irrational values while staying in the spin liquid phase~\cite{flores-calderon_irrational_2024}. However, such a tunable moment was found in {\it gapless} rank-1 and rank-2 $U(1)$ spin liquids. We expect that for a gapped fragile phase such as the one studied here the value of the fractional magnetic moment may change but cannot be deformed to a trivial value without closing the gap.

 \begin{figure}[ht]
    \centering
    \includegraphics[width=1\columnwidth]{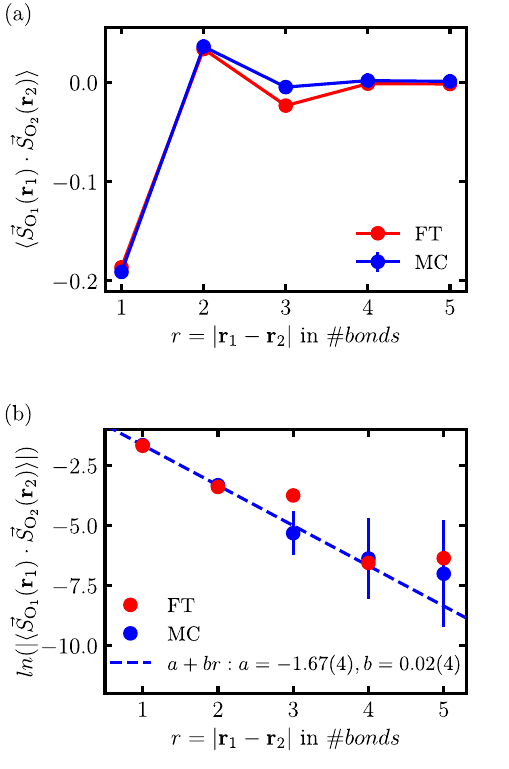}
    \caption{Orphan spin correlations as a function of the number of bonds connecting two dilution clusters for systems containing $L^3=10^3$ (FT) and $12\times4^3$ (MC) spins at temperature $\beta=1000$. (a) Orphan spin correlation function from FT and MC. (b) Logarithm of the absolute value of the orphan spin correlation function from MC and FT. The dashed-blue line corresponds to the linear fit of the MC data. The close agreement between data and fit indicates that the orphan spin correlations decay exponentially.}  
    \label{fig:SpinCorr}
\end{figure}

\subsection{Orphan spin correlations}
Next, we consider the case of two vacancy clusters each containing an orphan spin, $\vec{S}_{\text{O}_1}(\mathbf{r}_1)$ and $\vec{S}_{\text{O}_2}(\mathbf{r}_2)$ at positions $\mathbf{r}_1 $ and $\mathbf{r}_2$. We calculate the spin-spin correlations between the orphan spins as a function of their distance. As shown in Fig.~\ref{fig:SpinCorr}, we find from both, FT and MC an exponential decay in the correlations 
\begin{align}
    \expval{\vec{S}_{\text{O}_1}(\mathbf{r}_1)\cdot \vec{S}_{\text{O}_2}(\mathbf{r}_2)}\propto e^{-\abs{\mathbf{r}_1-\mathbf{r}_2}/\xi},
\end{align}
where the distance $\abs{\mathbf{r}_1-\mathbf{r}_2}$ is defined as the number of bonds between the two orphans. This is highly unusual, since for typical Coulomb spin liquids we would obtain in 3D a $1/\abs{\mathbf{r}_1-\mathbf{r}_2}$ decay characteristic of the charges being interpreted as charges of the emergent gauge field  \cite{moessner_magnetic_1999,moessner_pyrochlore_1998,IsakovSohndiSpinIce,sen_fractional_2011,sen_topological_2015,sen_vacancy-induced_2012,rehn_classical_2016}. Intuitively, we can understand the exponential decay from the fact that the $FT$-constrainers, $\textbf{T}_I(\textbf{q})= \sum_{\textbf{r}}\mathbf{C}_I(\mathbf{R}, \mathbf{r}) e^{i\textbf{q}\cdot \textbf{r}} $, are non-zero across the Brillouin zone, since the system is gapped. An expansion around the high symmetry point $\Gamma$ leads to a mass term at zeroth order $T^{\alpha}_I(\textbf{q}=0)=\sum_{i\in \alpha,i\in I}$ of order one. This means that the correlation length $\xi$ is also of order one in units of the lattice spacing, and indeed in our MC simulations we find that correlations in most directions are almost zero for distances beyond one unit cell.

\section{Ising and Quantum dimer model} \label{Sec:IsingandQuant}

 \begin{figure*}[ht!]
    \centering
    \includegraphics[width=0.98\columnwidth]{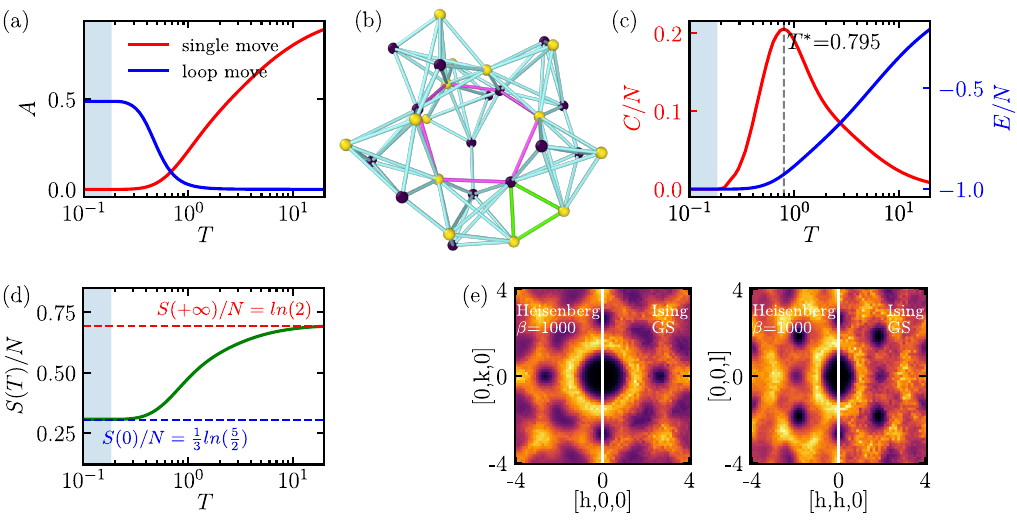}
    \caption{(a) Acceptance rate $A$ of single spin flips (red) and loop moves (blue) of the MC simulations for the Ising system as a function of temperature. The blue-shaded area indicates the range of temperatures for which the acceptance rate of the single move vanishes. The acceptance rate for the loop moves also takes into account the probability of successfully forming such loops without visiting a defect octahedron. Since defect octahedra proliferate at large temperatures and loops cannot be formed, the acceptance rate for loop moves vanishes in that limit. (b) Exemplary ground state configuration of the Ising model where the yellow and black dot represent up and down spins respectively. The bonds forming the shortest even length loop (six spins) are colored pink. The bonds forming the shortest loop (three spins) are colored in green. Flipping all the spins belonging to an even loop generates a new ground-state configuration. (c) Specific heat and energy per spin as a function of temperature for the Ising model. (d) Entropy per site obtained by numerical integration of the specific heat from the MC simulations. The dashed-blue line corresponds to the estimate of the entropy at zero temperature given by the Pauling counting argument. The dashed-red line corresponds to the entropy at infinite temperature, obtained by counting all possible states. (e) Structure factor comparison between the Ising and Heisenberg models, obtained with MC simulations. The structure factor of the Heisenberg model is computed at the lowest simulated temperature, while the one of the Ising model is computed in the ground-state manifold.}
    \label{fig:Ising}
 \end{figure*}
 
Let us briefly discuss the fate of our model for discrete Ising (rather than continuous Heisenberg) spins. We are interested in this case since it is the simplest limit where perturbations can be understood within a fully quantum mechanical description. The Ising Hamiltonian with discrete Ising spins $S_i^z$ for each point of the trilline lattice can be written in the same way as in Eq.~\eqref{Ham}. The ground states of this Hamiltonian are configurations where the total spin in each distorted octahedron is zero, which means that three spins point up and three spins point down. There are twenty possible configurations per octahedron satisfying this ice rule, which suggests an extensively degenerate ground state. Let us show this more explicitly within the independent octahedra (Pauling) approximation. A system with $N$ Ising spins can be in $2^N$ different configurations. In each octahedron there are ${6 \choose 3}=20$ ground states out of the $2^6$ possible spin configuations. Each octahedron has six spins which are all shared between two octahedra such that the total number of octahedra is $N/3$ leading to $2^N\times \left(\frac{20}{2^6}\right)^{N/3}=\left(\frac{5}{2}\right)^{N/3}$ ground state configurations. The expected entropy at zero temperature is then $S=\frac{N}{3}\ln(\frac{5}{2})\approx 0.3\ N$, about 50\% more than the residual entropy of spin ice on the pyrochlore lattice of corner-sharing tetrahedra. 
    
We confirm this estimate with MC simulations consisting of single-flip moves and the so-called short-loop moves, described in Ref.~\cite{Melko_2004_MCspinice}. In brief, a short-loop move consists of randomly generating strings of spins of alternating signs along the lattice bonds. A loop is closed when an octahedron is revisited, while it is discarded if it encounters a defective octahedron in which the spins do not sum to zero. Flipping the spins belonging to an even loop length leaves all visited octahedra in a ground state configuration. The shortest even loop, illustrated in Fig.~\ref{fig:Ising}(b), is formed by six spins. As shown in Fig.~\ref{fig:Ising}(a), the acceptance rate of single-flip moves drops to zero as the system reaches the ground state, while the acceptance rate for loop moves stabilizes at around 0.5, as only the loops with an even number of spins are accepted. To increase the acceptance rate of the loop moves, the algorithm excludes the three-spin loop also shown in Fig.~\ref{fig:Ising}(b). 
The specific heat in Fig.~\ref{fig:Ising}(c) features the characteristic bump of a disordered system connected to the high-temperature paramagnetic state. Fig.~\ref{fig:Ising}(d) shows the entropy computed by numerically integrating the approximated thermodynamic relation
\begin{align}
    S(T)&=S(+\infty)-\int_{T}^{+\infty}\frac{C(T^\prime )}{T^\prime}dT^\prime \notag\\
    &\approx N\ln(2)-\int_{T}^{T_{\rm max}}\frac{C(T^\prime )}{T^\prime}dT^\prime,
\end{align}
where $C$ is the specific heat obtained from the MC simulations. At low temperatures, the entropy agrees with the simple argument above. Furthermore, the static spin structure factor in Fig.~\ref{fig:Ising}(e) shows a pattern similar to that found for the Heisenberg case and the prediction from the SCGA calculation. We may thus conclude from this characterization that the Ising model does not order but possibly realizes a CSL state.

For a deeper theoretical understanding of the Ising case and later the quantum model, we may adopt an alternative perspective where instead of Ising spins on the trilline lattice, we identify the degrees of freedom as Ising variables at the midpoints of the bonds connecting the sites of the dual lattice, the trillium lattice. Thus, the system can be viewed as a dimer model on the trillium lattice where an up-spin corresponds to a dimer placed on that bond while a down-spin corresponds to no dimer. The ground state condition of the trilline Ising model (corresponding to three spins up and three spins down in each octahedron) then translates into the condition of three dimers emanating from each trillium lattice site, so-called triple dimer coverings. Crucially, the trillium lattice is {\it non-bipartite}. It has previously been noted that dimer coverings on non-bipartite 2D and 3D lattices can be distinguished by a $\mathbb{Z}_2$ winding parity~\cite{rehn_fractionalized_2017,moessner_three-dimensional_2003,huse_coulomb_2003}. The winding parity for 3D models like our trillium system can be explained in analogy to fluxes through a reference surface. 

Let us consider a system with periodic boundaries along the $x$, $y$, and $z$ directions. Furthermore, we choose a unit vector $\vec{e}_{\mu}$ ($\mu=x,y,z$) along one of these directions and a reference plane perpendicular to $\vec{e}_{\mu}$ which we denote $\Sigma_\mu$ and which extends over the full periodic system. We assume that the plane cuts $N_\mu$ dimers. Next, we consider a local move that keeps the system within the ground state manifold. Such a move corresponds to switching the dimer occupation on a contractable loop of alternating occupation. If the loop does not pierce $\Sigma_\mu$, the number of cut dimers $N_\mu$ remains unchained, otherwise it will necessarily pierce the plane twice. The loop move may change the number of cut dimers $N_\mu$ but the parity $N_\mu(\text{mod }2)$ remains unchanged, establishing a $\mathbb{Z}_2$ topological index associated with the plane $\Sigma_\mu$, or more generally, associated with the direction $\vec{e}_{\mu}$. The argument is analogous for the other two spatial directions giving rise to $\mathbb{Z}_2$ flux sectors that establish a topological magnet. Thus our Ising model corresponds to a $\mathbb{Z}_2$ CSL in 3D.

We may now upgrade our discussion to the quantum level by considering perturbations on top of the Ising ground states within degenerate perturbation theory. Such perturbations correspond to off-diagonal terms in the original $S_i^z$-basis such as transverse fields $\Gamma \sum_i S^x_i$ or $XY$ interaction terms $J_{\perp} \sum_{\langle i,j\rangle} (S_i^{+}S_j^{-}+\text{h.c.})$ which flip spins and generically take the system out of the Ising ground state manifold. Assuming $J \gg J_{\perp},\Gamma$ such that defect octahedra have a large energy cost, the terms in the perturbation series are given by processes that lead the system from one ground state to the other. These processes, so-called loop resonances correspond to the aforementioned flips of alternating spins on even length loops. In lowest order these are the six-site resonances illustrated in Fig.~\ref{fig:Ising}(b) and in the language of quantum dimer models we can write these perturbations as
\begin{align}
    &H_{\text{QDM}}=\sum_{\hexagon}\left\{ -t \left( \ket{\raisebox{-0.4\height}{
\includegraphics[width=1cm]
{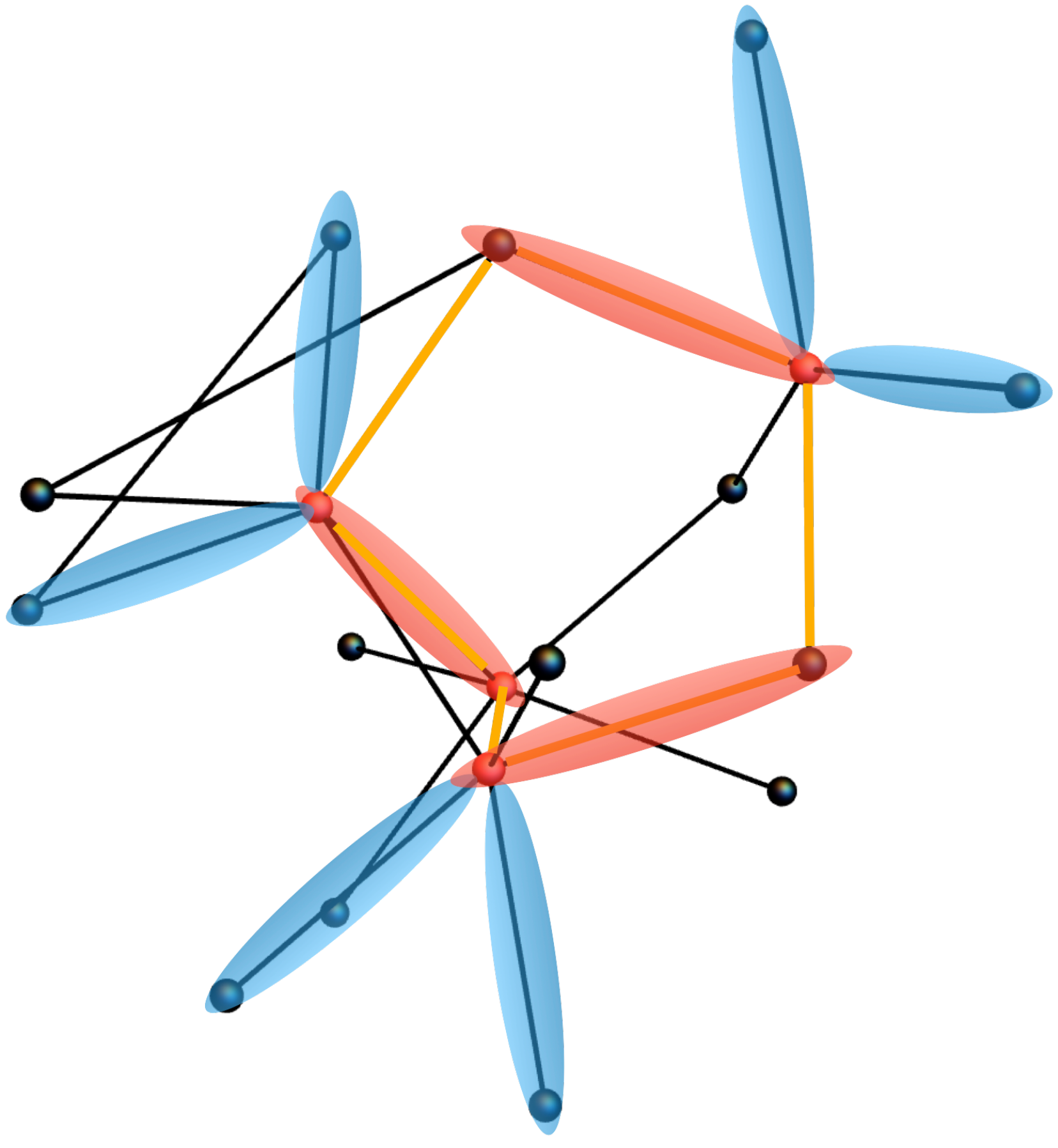}}}\bra{\raisebox{-0.4\height}{
\includegraphics[width=1cm]
{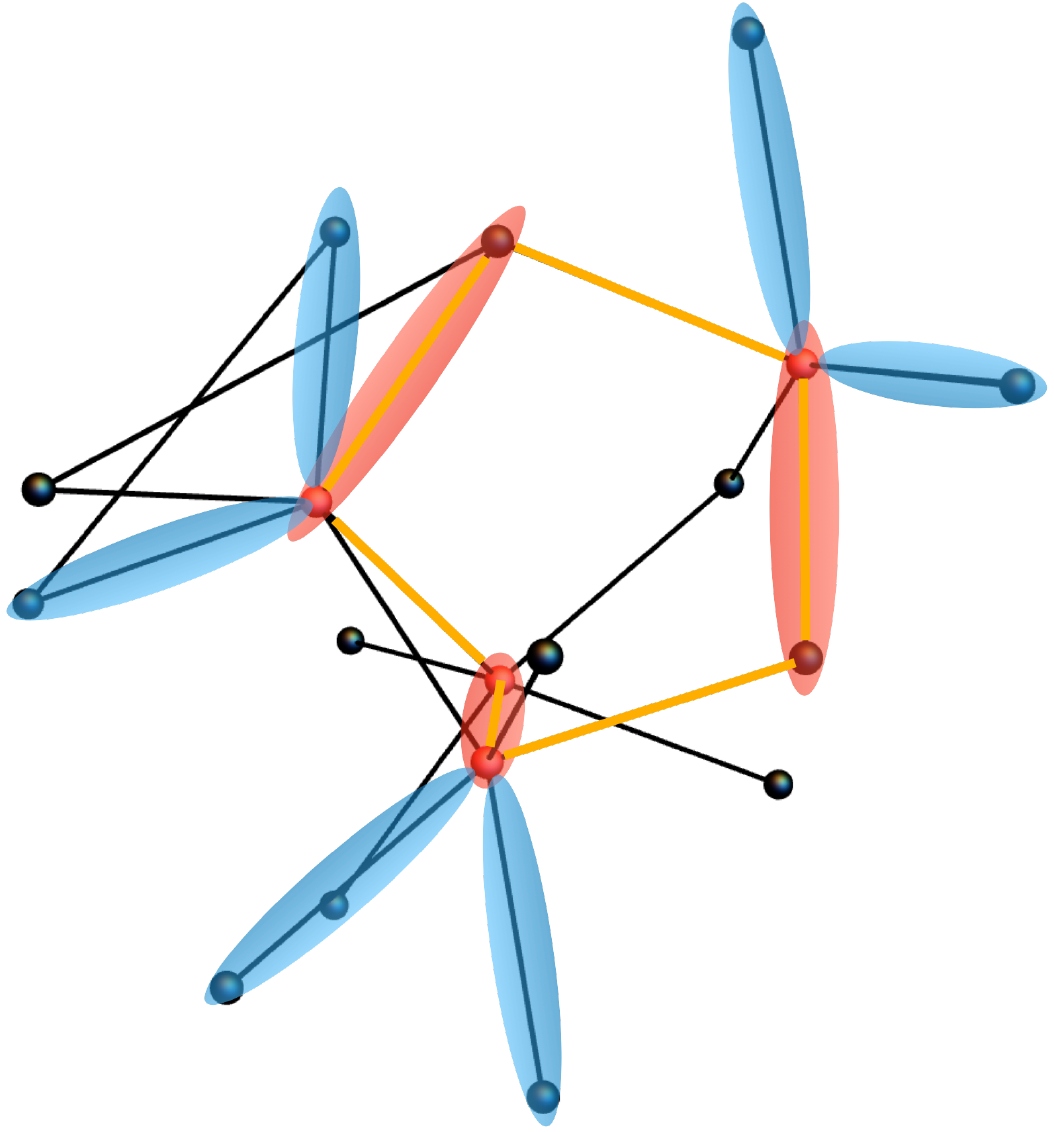}}}+\text{h.c.}\right)\notag\right.\\
&\left.+v \left( \ket{\raisebox{-0.4\height}{
\includegraphics[width=1cm]
{T1Trillium.pdf}}}\bra{\raisebox{-0.4\height}{
\includegraphics[width=1cm]
{T1Trillium.pdf}}}+\ket{\raisebox{-0.4\height}{
\includegraphics[width=1cm]
{T2Trillium.pdf}}}\bra{\raisebox{-0.4\height}{
\includegraphics[width=1cm]
{T2Trillium.pdf}}}\right)\right\},
\end{align}
where we sum over all six-bond loops $\hexagon$ of the trillium lattice. Following Ref.~\cite{hermele_pyrochlore_2004} the resonance is of order $t\propto \Gamma^6/J^5$ or $-J_{\perp}^3/J^2$, where for later arguments we require a ferromagnetic $J_\perp<0$. Furthermore, we have added a diagonal term with strength $v$ \cite{RKpoint,anderson_resonating_1973,triangdimer,moessner_TriangRes_2001} which counts the number of \textit{flippable} loops in a given state. For negative $v\rightarrow -\infty$ this term generates a ground state with the maximal number of flippable loops, referred to as a columnar state. The opposite limit $v\rightarrow \infty$ is characterized by no flippable loops and is generically called a staggered state. The phase diagram as a function of $v/t$ is then generically divided into three regions. From a spin liquid perspective, the center region at $v/t \approx 1$ is the most interesting regime since it can give rise to emergent deconfined $U(1)$ or $\mathbb{Z}_2$ gauge theories \cite{DimerModelLiquids}. 

Of particular importance is the RK point at $v=t$~\cite{RKpoint}, where we can rewrite the Hamiltonian as a sum of projectors:
\begin{align}\label{eq:rk}
    H_{\text{RK}}= t \sum_{\hexagon} \ket{\Phi_{\hexagon}}\bra{\Phi_{\hexagon}},\ \ \ket{\Phi_{\hexagon}}=\ket{\raisebox{-0.4\height}{
\includegraphics[width=1cm]
{T1Trillium.pdf}}}-\ket{\raisebox{-0.4\height}{
\includegraphics[width=1cm]
{T2Trillium.pdf}}}.
\end{align}
For positive $t$ all eigenvalues are non-negative. Let us consider the classical Ising configurations which, as described above, can be grouped into $\mathbb{Z}_2$ flux sectors. An application of the resonance term connects states only within each sector as the contractable loops are local moves. This allows us to define a state $\ket{\Psi}$ as an equal weight superposition of all triple dimer coverings in the trillium lattice for each $\mathbb{Z}_2$ sector. Since $\ket{\Psi}$ is annihilated by each of the projectors $\ket{\Phi_{\hexagon}}\bra{\Phi_{\hexagon}}$ in Eq.~(\ref{eq:rk}) the wave function $\ket{\Psi}$ is the exact and unique ground state in a given $\mathbb{Z}_2$ sector at the RK point. This argument assumes that all states in a $\mathbb{Z}_2$ sector are connected by resonance loops, which, however, is not always the case. For example, the dimer model on the triangular lattice contains staggered states disconnected from any other states~\cite{moessner_TriangRes_2001}. In such cases only the connected states contribute to the equal weight superposition in $\ket{\Psi}$ and different degenerate ground states in a $\mathbb{Z}_2$ sector exist. Since the minimal six-bond resonance loops of our trillium system are relatively short (of the same length as in pyrochlore spin ice) flippable loops are abundant such that one can expect large connected sectors with spin liquid behavior. 

Moreover, equal-time correlations are accessible at the RK point, which for an observable diagonal in the dimer basis can be calculated by an equal weight summation over all triple dimer coverings, the same procedure as for the calculation of zero-temperature correlations in the pure Ising model. Thus our zero-temperature MC results for the Ising model also hold for the quantum model at the RK point and we can conclude that the latter system also has short range and exponentially decaying correlations. Furthermore, physical configurations are distinguished by their $\mathbb{Z}_2$ winding parity characterizing the resulting state as a $\mathbb{Z}_2$ quantum spin liquid. The stability region of such a state is not clear from the above arguments, but following previous results for 3D dimer models in non-bipartite lattices a finite quantum spin liquid region in the phase diagram is possible~\cite{DimerModelLiquids}. 

To summarize, we have investigated the classical Ising model on the trilline lattice, derived from a fragile spin liquid model. We showed an absence of long-range correlations and the emergence of an extensively degenerate ground state manifold characterized by a $\mathbb{Z}_2$ winding parity. Quantum mechanical perturbations on top of the ground state can be analyzed within in the framework of quantum dimer models, for which we showed the emergence of a $\mathbb{Z}_2$ quantum spin liquid at the RK point.

\section{Conclusions}\label{sec:conclusion}

We have constructed a classical Heisenberg model on the line graph of the trillium lattice, the trilline lattice, which has no order down to the lowest temperatures. We find exponentially decaying spin-spin correlations which manifest themselves as a broad continuum in the spin structure factor. Moreover, we showed that such a state has a fractionalized magnetic response for a diluted system with non-magnetic vacancies. Specifically, we find that the relevant vacancy clusters realize an effective $S/2$ free spin, characterizing the system as distinctly different from a trivial paramagnet. Nevertheless, we confirm by FT and MC calculations that such fractionalized charges are also short-range correlated with a characteristic exponential decay. This is in contrast to usual Coulomb spin liquids which exhibit power-law correlations characteristic of the emergent gauge charges, for example the monopoles of spin ice. Examining the Ising case we again find an absence of ordering and an extensive ground state degeneracy. Furthermore, by mapping the Ising model to triple dimer coverings on the trillium lattice we establish the existence of $\mathbb{Z}_2$ flux sectors and find a $\mathbb{Z}_2$ quantum spin liquid state at the RK point of the system. We conclude from these results, that our model realizes the first instance of a fragile spin liquid in three dimensions.\\

\textit{Acknowledgments}--- We are grateful to Benoit Doucot for fruitful discussions. This work was in part supported by the Deutsche
Forschungsgemeinschaft under Grants No. SFB 1143
(Project No. 247310070) and the cluster of excellence
ct.qmat (EXC 2147, Project No. 390858490). J.~R.~acknowledges support from the Deutsche Forschungsgemeinschaft within Project-ID 277101999 CRC 183 (Project A04). The work of J.~R. was performed, in part, at the Aspen Center for Physics, which is supported by National Science Foundation Grant No.~PHY-2210452. The MC simulations were performed on the Sheldon cluster at the Physics Department of Freie Universit{\"a}t Berlin.

\appendix

\section{Details on the analytical calculations}\label{app_a}
The calculations for the Heisenberg model of Eq.~\eqref{Ham} were carried out within the framework of the self-consistent Gaussian approximation (SCGA) and vacancy FT. For this, we followed Refs.~\cite{garanin_classical_1999,IsakovSohndiSpinIce,sen_vacancy-induced_2012,rehn_classical_2016,IsakovSondhiLargeN,flores-calderon_irrational_2024}. In particular, the calculations for the orphan spin magnetic moment and the spatial correlations are concisely shown 
in the Supplementary Material of Ref.~\cite{flores-calderon_irrational_2024}. 

\section{Details on the Monte Carlo simulations}\label{app_b}
We simulate a system of $N=12 \times L \times L \times L$ spins, with periodic boundary conditions. The spins are initialized  in a random configuration and gradually cooled down. For the MC simulations of the Heisenberg model, the initial temperature is set to $T=10$, as in the main text $T$ is expressed in units of $J$ with $J=1$. A single MC move consists of $N$ heat bath moves~\cite{Miyatake1986heatbath} each followed by 10 overrelaxation moves~\cite{Creutz1987overrelax}. The results in Fig.~\ref{fig:SSF}(c), (d) and Fig.~\ref{fig:Mag} are obtained by simulating a linear size $L=8$ and averaging over 100 independent runs. The results in Fig. \ref{fig:SpinCorr} are obtained by simulating a linear size $L=4$ and averaging over 500 independent runs. The error bars correspond to the standard deviation across the runs.\\

For the  MC simulations of the Ising model, the initial temperature is set to $T=20$ and a  single MC move consists of $N$ single spin flip moves and $N$ loop moves, described in Sec.~\ref{Sec:IsingandQuant}. Both types of moves are accepted or rejected according to the Metropolis scheme. When reaching the temperature at which single flip moves are always rejected, only loop moves are applied. The results in Fig.~\ref{fig:Ising} are obtained by simulating a linear size $L=8$ and averaging over 100 independent runs.

\bibliography{main.bib}

\begin{thebibliography}{57}
\bibitem [{foo()}]{footnote}%
  \BibitemOpen
  \href@noop {} {}\bibinfo {note} {For simplicity we omit the vector notation for the $\mathcal{N}$-component spins in this section. Within this constraint formalism the different spin components can be viewed as independent degrees of freedom such that a distinction between different components is not necessary.}\BibitemShut {Stop}%
\end{thebibliography}

\clearpage

\end{document}